\documentclass[twocolumn,epjc3]{svjour3} 
\smartqed
\usepackage[T1]{fontenc}
\usepackage{epsfig,amsmath,graphicx,multirow,amssymb}
\usepackage[colorlinks,citecolor=blue,urlcolor=blue,linkcolor=blue]{hyperref}
\journalname{Eur. Phys. J. C}

\newcommand{\ie}{{\it i.e.}}

\newcommand{\be}{\begin{equation}}

\newcommand{\ee}{\end{equation}}
\def\bsp#1\esp{\begin{split}#1\end{split}}
\def\bpm{\begin{pmatrix}}
\def\epm{\end{pmatrix}}

\newcommand{\resum}{{\sc Resummino}}
\newcommand{\cpp}{{\sc C++}}
\newcommand{\slhaea}{{\sc SLHAea}}
\newcommand{\fortran}{{\sc Fortran}}
\newcommand{\boost}{{\sc Boost}}
\newcommand{\lhapdf}{{\sc LhaPdf}}
\newcommand{\gsl}{{\sc GSL}}
\def\d{{\rm d}}
\def\eps{\epsilon}

\begin{document}

\title{Precision predictions for electroweak superpartner production at hadron colliders with
  \resum}
\author{
  Benjamin Fuks\thanksref{a,b} \and
  Michael Klasen\thanksref{c} \and
  David R.\ Lamprea\thanksref{c} \and
  Marcel Rothering\thanksref{c}
}
\institute{Theory Division, Physics Department, CERN, CH-1211 Geneva 23, Switzerland\label{a}
\and
  Institut Pluridisciplinaire Hubert Curien/D\'epartement Recherches
  Subatomiques, Universit\'e de Strasbourg/CNRS-IN2P3, 23 Rue du Loess, F-67037
  Strasbourg, France\label{b}
\and
  Institut f\"ur Theoretische Physik, Westf\"alische Wilhelms-Universit\"at M\"unster,
  Wilhelm-Klemm-Stra\ss{}e 9, D-48149 M\"unster, Germany\label{c}
}
\date{Received: date / Accepted: date}

\maketitle

\vspace*{-8cm} \noindent 
\small{CERN-PH-TH/2013-064, IPHC-PHENO-13-02, MS-TP-13-06}
\vspace*{69mm}

\begin{abstract}
  We describe the \resum\ package, a \cpp\ and \fortran\ program dedicated to precision
  calculations in the fra\-me\-work of gaugino and slepton pair production at hadron colliders.
  This code allows
  to calculate transverse-momentum and invariant-mass distributions as well as total cross
  sections by combining the next-to-leading order predictions obtained by means of perturbative
  QCD with the resummation of the large logarithmic contributions arising in the small
  transverse-momentum region and close to the production threshold. The results computed in
  this way benefit from reduced theoretical uncertainties, compared to a pure next-to-leading
  order approach as currently employed in the experimental analyses searching for sleptons and
  gauginos at hadron colliders. This is illustrated by using of \resum\ in the context of a
  typical supersymmetric benchmark point dedicated to superpartner searches at the Large Hadron
  Collider.
\end{abstract}


\section{Introduction}
\label{sec:intro}

Over the last forty years, theoretical developments and experimental discoveries in high-energy
physics have led to an extremely coherent picture, the so-called Standard Model of
particle physics. In particular, the recent observation of a neutral bosonic particle
compatible with a Standard-Model-like Higgs boson \cite{Aad:2012gk,Chatrchyan:2012gu}
represents an impressive success of this theoretical framework. However, many questions, such
as the stabilization of the mass of a fundamental scalar particle with respect to radiative
corrections, remain unanswered. Consequently, the Standard Model is widely accepted as an
effective theory implied by a more fundamental one. Among the plethora of candidates for this
new physics theory, weak-scale supersymmetry \cite{Nilles:1983ge,Haber:1984rc} is one of the
most appealing and studied option. It not only addresses the above-mentioned hierarchy
problem but
also provides a solution for the unification of the gauge couplings at high energies and
explains the presence of dark matter in the Universe.

Experimental searches, especially at the Large Hadron Collider (LHC) at CERN, for the
supersymmetric partners of the Standard Model particles are therefore among the main items
of the experimental program in high-energy physics. Up to now, both the ATLAS and CMS
collaborations have mainly focused on seeking for hints of squarks and gluino, being strongly
produced. All results are however compatible with the Standard Model 
expectation~\cite{atlassusy,%
cmssusy}. As a consequence, limits on the masses of the squarks and gluino are pushed to higher
and higher scales and
the experimental attention starts to shift towards the pair production of the electroweak
slepton, neutralino and chargino eigenstates. 

Investigations at a center-of-mass energy of
8~TeV of the trilepton golden signature have already led to bounds of several hundreds of GeV 
on the masses of these particles~\cite{ATLAS-CONF-2012-154,CMS-PAS-SUS-12-022}. However,
in contrast
to the strong production channels where estimates of signal cross sections rely on precise
theoretical predictions at the next-to-leading order and next-to-leading logarithmic accuracy
\cite{Kulesza:2008jb,Kulesza:2009kq,%
Beenakker:2009ha,Beenakker:2010nq,Beenakker:2011fu,Kramer:2012bx,Falgari:2012hx}, searches
for the weak superpartners are only based on next-to-leading order computations, suffering
from larger theoretical uncertainties~\cite{Beenakker:1999xh,Bozzi:2007qr,Debove:2010kf}.

For an efficient suppression of the Standard Model background and a more precise extraction
of the
underlying supersymmetric mass limits (or measurement of the supersymmetric parameters in the
case of a discovery), accurate theoretical calculations of signal cross
sections and key kinematical distributions are imperative. Along these lines, while
supersymmetric particle pairs are produced with a vanishing transverse momentum ($p_T$) at the
lowest order in the strong coupling $\alpha_s$, gluon radiation from quark-antiquark initial
states and their splitting into quark-antiquark pairs at ${\cal O}(\alpha_s)$ induce transverse
momenta extending to larger values. Next-to-leading order corrections have therefore to be
taken into account.
However, the perturbative calculation diverges at small $p_T$, which indicates the need for
soft-gluon resummation to all orders and for a consistent matching with the
fixed order results. On the same footings, when supersymmetric particle pairs are produced
close
to the production threshold, soft gluon emission again leads to large logarithmic terms that
must be resummed and matched to fixed order in order to obtain reliable predictions.

These considerations motivate us to introduce in this work the computer code \resum,
implemented in \cpp\ and \fortran\ and that can be downloaded from the website

\verb+http://www.resummino.org+

It combines a leading order calculation of slepton and gaugino pair production at hadron
colliders~\cite{Bozzi:2004qq,Bozzi:2007me,Debove:2008nr} supplemented by the associated
supersymmetric QCD next-to-leading order corrections \cite{Bozzi:2007qr,Debove:2010kf},
with the resummation of the leading and next-to-leading
logarithms to all orders in the threshold \cite{Bozzi:2007qr,Debove:2010kf,Fuks:2012qx}
and small transverse-momentum
\cite{Fuks:2012qx,Bozzi:2006fw,Debove:2009ia} regimes or simultaneously in
both~\cite{Fuks:2012qx,Bozzi:2007tea,Debove:2011xj}.

In Section~\ref{sec:formalism} of this paper, we briefly review the three resummation
formalisms included in the \resum\ package, giving the explicit form of the implemented
formulas, as well as the techniques employed to calculate the underlying integrals.
Section~\ref{sec:resum} is more technical and is dedicated to the installation of the program
and its running. A set of illustrative results are then shown in
Section~\ref{sec:benchmarking},
after choosing a benchmark point typical for supersymmetry searches at the LHC.

\section{Resummation formalisms}\label{sec:formalism}

\subsection{Main features}

In perturbative QCD, the doubly-differential cross section describing the production,
at hadron colliders, of a gaugino or a slepton pair with an invariant mass $M$ and a transverse
momentum $p_T$ is calculated by means of the QCD factorization theorem. The fixed order result,
is in this way obtained by convolving the partonic cross section
$\d\sigma_{ab}$, computed at a given order in the strong coupling, with the universal
densities $f_a$ and $f_b$ of the
partons $a$ and $b$ carrying the momentum fractions $x_a$ and $x_b$ of the colliding hadrons,
\be\label{eq:d2sigX}\bsp
  & M^2 \frac{\d^2 \sigma}{\d M^2 \d p_T^2}(\tau) = \sum_{ab}
    \int_0^1 \! \d x_a \d x_b \d z \Big[x_a f_a(x_a,\mu_F^2) \Big]
\\&\ \times\!
    \Big[x_b f_b(x_b, \mu_F^2)\Big]
    \Big[z \d\sigma_{ab}(z,M^2,p_T^2,\mu_F^2,\mu_R^2)\Big] \delta(\tau \!-\! x_ax_bz) \ .
\esp\ee
In this expression, the unphysical factorization and renormalization scales are respectively
denoted by $\mu_F$ and $\mu_R$ and we have introduced the quantity $\tau=M^2/S_h$, $S_h$
standing for the hadronic center-of-mass energy.
After performing a Mellin transform with respect to the variable $\tau$, this cross section
can be reexpressed as a simple product of parton densities and the partonic cross section
in the conjugate Mellin $N$-space,
\be\bsp
  & M^2 \frac{\d^2\sigma}{\d M^2 \d p_T^2}(N-1)=\sum_{ab} f_a(N,\mu_F^2)
    f_b(N,\mu_F^2)
\\ & \qquad \qquad \times  \sigma_{ab}(N,M^2,p_T^2,\mu_F^2, \mu_R^2)\ ,
\esp\label{eq:d2sig}\ee
where the Mellin moments of the quantities  $F = \sigma$, $\sigma_{ab}$, $f_a$ and $f_b$
are defined by
\be
  F(N) = 
    \int_0^1 \d y\ y^{N-1}\ F(y)  \ ,
\ee
with $y = \tau$, $z$, $x_a$ and $x_b$, respectively.
Under the form of Eq.~\eqref{eq:d2sig},
it is possible to effectively resum to all orders in $\alpha_s$ the large
logarithmic terms arising in the critical regions, \ie, when the transverse momentum tends
towards zero and/or when the partonic center-of-mass energy is close to the production
threshold. The exact form of the resummed cross sections is detailed, for the small transverse
momentum, threshold and joint regimes in Section~\ref{sec:thresh}, Section~\ref{sec:pt} and
Section~\ref{sec:joint}, respectively.

Although these large logarithms must clearly be resummed in the critical regions, the full
perturbative computation, only partially accounted for by resummation, is expected to be
reliable
away from these regions. In order to obtain valid predictions in all kinematical regions, the
fixed order ($\sigma^{\rm (f.o.)}$) and resummed ($\sigma^{\rm (res.)}$) results have then to
be consistently combined by subtracting from their sum their overlap ($\sigma^{\rm (exp.)}$),
\be\label{eq:mtch}
 \sigma_{ab}=
 \sigma^{\rm(res.)}_{ab}+\sigma^{\rm(f.o.)}_{ab}-\sigma^{\rm(exp.)}_{ab} \ .
\ee
The latter is obtained by expanding the resummation formula to the desired accuracy, \ie, at
${\cal O}(\alpha_s)$ in our case, and is thus dependent on the employed resummation formalism.
The analytical expression of $\sigma^{\rm (exp.)}$ are therefore given in the relevant
subsection  below.

While the fixed order result $\sigma^{\rm (f.o.)}$ can in general be directly computed in
physical space, or in other words by directly using Eq.~\eqref{eq:d2sigX}, the resummed
component and the calculation of its expansion at a given order in $\alpha_s$ requires
factorization properties that only hold in conjugate
spaces so that Eq.~\eqref{eq:d2sig} is employed. An inverse Mellin transform is consequently
necessary to get back to the physical space, which implies to take care of the singularities
possibly arising at the level of the $N$-space cross section. This is achieved by choosing an
integration contour inspired by the principal value procedure and minimal
prescription~\cite{Contopanagos:1993yq,Catani:1996yz}. On this contour, the Mellin variable $N$
is parameterized as a function of two parameters $C$ and $\phi$ and one variable~$y$,
\be
  N(y) = C + y e^{\pm i\phi} \quad\text{with}\quad y\in[0,\infty[ \ .
\ee
The parameter $C$ is chosen such that, on the one hand, the poles in the Mellin moments of the
parton densities related to the Regge singularity lie to the left of the integration
contour and that, on the other hand, the Landau pole related to the running of the strong
coupling constant lies to its right.
Furthermore, the phase $\phi$ can formally be chosen anywhere in the range~$[\pi/2,\pi[$.

\subsection{Threshold resummation at the next-to-leading logarithmic accuracy}
\label{sec:thresh}
In the threshold regime, the slepton or gaugino pair invariant mass $M^2$ is close to
the partonic center-of-mass energy or equivalently, the Mellin variable $N$ tends
towards infinity. In this case, refactorization allows us
to rewrite the partonic cross section $\sigma_{ab}$ obtained after integrating
Eq.~\eqref{eq:d2sig} over the transverse momentum $p_T$ into a closed
exponential form~\cite{Sterman:1986aj,Catani:1989ne,Catani:1990rp,Kidonakis:1997gm,%
Kidonakis:1998bk,Vogt:2000ci},
\be\bsp
  &\sigma_{ab}^{\rm (res.)}(N,M^2,\mu_F^2, \mu_R^2) =
  {\cal H}_{ab}(M^2, \mu_F^2,\mu_R^2)
\\ & \qquad \qquad \times \exp\Big[{\cal G}_{ab}(N,M^2,\mu_F^2,\mu_R^2)\Big] \ .
\esp\label{eq:resori}\ee
The hard part of the scattering process, independent of the Mellin variable $N$, is embedded
within the perturbatively computable function ${\cal H}_{ab}$. On different footings, the
function ${\cal G}_{ab}$, that can also be computed by means of perturbation theory, collects
soft and collinear parton
radiation and absorbs the large logarithmic contributions arising at fixed order.

It has however recently been shown that Eq.~\eqref{eq:resori} can be improved, at the
next-to-leading logarithmic accuracy, by including and resumming subleading terms stemming from
universal collinear radiation of the initial state partons~\cite{Kramer:1996iq,Catani:2001ic,%
Kulesza:2002rh,Almeida:2009jt}. This improvement procedure leads to a resummed cross section
explicitly depending on the one-loop approximation of the QCD evolution operator
$E_{ab}^{(1)}$, which drives the behavior of the parton-into-parton density functions with the
energy and encompasses collinear radiation \cite{Furmanski:1981cw}. Consequently, the original
resummation formula is modified to
\be\bsp
 & \sigma^{\rm (res.)}_{ab}(N,M^2,\mu_F^2, \mu_R^2) =  \sum_{a',b'}
    E_{aa'}^{(1)}(N,M^2/\bar{N}^2,\mu^2_F)\
\\ &\quad
    \times
    E_{bb'}^{(1)}(N,M^2/\bar{N}^2,\mu^2_F)
    \tilde{\cal H}_{a'b'}(M^2,\mu^2_R)
\\&\quad
    \times
    \exp\Big[\tilde{\cal G}_{a'b'}(\bar N,M^2,\mu^2_R)\Big] \ ,
\esp\label{eq:resth}\ee
where we have introduced the reduced Mellin variable defined by $\bar{N} = N e^{\gamma_E}$.
The improved soft and hard functions $\tilde{\cal G}_{ab}$ and $\tilde{\cal H}_{ab}$
can still be perturbatively computed and read, at the next-to-leading logarithmic accuracy,
\be\bsp
  \tilde{\cal H}_{ab}(M^2,\mu_R^2) = &\
    \tilde{\cal H}_{ab}^{(0)}(M^2, \mu_R^2) \!+\!
    \frac{\alpha_s(\mu_R^2)}{2 \pi} \tilde{\cal H}_{ab}^{(1)}(M^2,\mu_R^2) \ , \\
  \tilde{\cal G}_{ab}(N,M^2,\mu_R^2) = &\ \tilde{g}_{ab}^{(1)} \Big(\frac{\alpha_s}{2 \pi}
    \beta_0 \ln \bar{N}\Big) \ \ln\bar N 
\\ &\qquad +
    \tilde{g}_{ab}^{(2)} \Big(\frac{\alpha_s}{2 \pi} \beta_0 \ln \bar{N}, \frac{M^2}{\mu_R^2}
    \Big) \ .
\esp\ee
The arguments of the leading and next-to-leading contributions to the Sudakov form factor
$\tilde{\cal G}_{ab}$ depend, in addition to the reduced Mellin variable, on the one-loop
coefficient of the QCD beta-function $\beta_0$.
This quantity and the associated two-loop coefficient $\beta_1$ given for further references
are defined, in our normalization conventions, by
\be\bsp
  \beta_0 =&\ \frac{11}{6} C_A - \frac23 N_f \tau_R \ , \\
  \beta_1 =&\ \frac16 \Big[17 C_A^2 - 5C_A N_f - 3 C_F N_f\Big] \ ,
\esp\ee
for $N_f$ active quark flavors. In those expressions, the group theory invariants of $SU(3)$
are standard and read $C_A = 3$, $C_F=4/3$ and $\tau_R=1/2$.

The first two coefficients $\tilde{g}^{(1)}_{ab}$ and $\tilde{g}^{(2)}_{ab}$ of the function
$\tilde{\cal G}_{ab}$ allow to resum the leading and next-to-leading logarithmic
contributions yielded by soft and collinear radiation. In the $\overline{\rm MS}$ 
renormalization scheme, their functional form is explicitly given
by~\cite{Bozzi:2007qr,Debove:2010kf,Sterman:1986aj,Catani:1989ne,Catani:1990rp,Kramer:1996iq}
\be\bsp
  & \tilde{g}_{ab}^{(1)}(\lambda)  = \frac{1}{2\lambda\beta_0}
    \Big(A_{a}^{(1)}+A_{b}^{(1)}\Big) \Big(2\lambda+\ln(1-2\lambda)\Big) \ , \\
  & \tilde{g}_{ab}^{(2)}\Big(\lambda,\frac{M^2}{\mu_R^2}\Big)=
     - \frac{1}{2 \beta_0^2} \Big[A_{a}^{(2)} \!+\!A_{b}^{(2)}\Big]
      \Big[2\lambda+\ln(1-2\lambda)\Big]\\
  &\ +\frac{1}{\beta_0} \Big[B_{a}^{(1)} \!+\! B_{b}^{(1)}\Big] \ln(1-2\lambda) \\
  &\ + \frac{1}{2\beta_0}\Big[A_{a}^{(1)} \!+\! A_{b}^{(1)}\Big]
    \Big[2\lambda+\ln(1-2\lambda)\Big] \ln\frac{M^2}{\mu_R^2} \\
  &\ + \frac{\beta_1}{2 \beta_0^3}\Big[A_{a}^{(1)} \!+\! A_{b}^{(1)}\Big] 
    \Big[2\lambda \!+\! \ln(1\!-\!2\lambda) \!+\! \frac{1}{2}\ln^2(1 \!-\! 2\lambda)\Big] \ .
\esp\ee
This shows that the first two terms of the perturbative expansion of the Sudakov form factor
only depend on the $A_a$ and $B_a$ resummation functions respectively describing soft and
collinear radiation and flavor-conserving collinear radiation. The leading terms of their
expansion as series in powers of $\alpha_s$ have been calculated in the past and are given
by~\cite{Catani:1989ne,Kodaira:1981nh,Catani:1988vd}
\be
  A_a^{(1)} = 2C_a \ , \qquad
  A_a^{(2)} = 2C_a\bigg[\bigg(\frac{67}{18}- \frac{\pi^2}{6}\bigg)C_A -\frac{5}{9}N_f\bigg] \ .
\label{eq:A1A2}\ee
and 
\be\label{eq:B1}
  B_q^{(1)} = -3 C_F \qquad\text{and}\qquad B_g^{(1)} = -2\beta_0 \ .
\ee
for the $A_a$ and $B_a$ functions.

Turning to the hard parts of the resummed cross section, the leading and next-to-leading
order contributions to the $\tilde{\cal H}_{ab}$ function read~\cite{Bozzi:2007qr,%
Debove:2010kf}
\be\bsp
  \tilde{\cal H}_{ab}^{(0)}(M^2,\mu_R^2) =&\ \sigma_{ab}^{(0)}(M^2) \ , \\
  \tilde{\cal H}_{ab}^{(1)}(M^2,\mu_R^2) =&\ \sigma_{ab}^{(0)}(M^2)
    \bigg[ {\cal A}_0 +  \frac{\pi^2}{6} \Big( A^{(1)}_a + A^{(1)}_b\Big) \bigg] \ .
\esp\label{eq:Hthre}\ee
where the dependence on the renormalization scale $\mu_R$ of the infrared-finite part
${\cal A}_0$ of the renormalized virtual corrections $\sigma_{ab}^V$ is understood. The
expression of the $\tilde{\cal H}_{ab}^{(1)}$ component is not general and assumes that
the virtual contributions to the next-to-leading order cross section are normalized, in
$4-2\eps$ dimensions, as
\be\label{eq:defa0}\bsp
  \sigma_{ab}^{V}(M^2,\mu^2_R) = & \frac{\alpha_s}{2 \pi} \bigg(\frac{4\pi\mu_R^2}{M^2}
  \bigg)^\eps \frac{\Gamma(1 - \eps)}{\Gamma(1 - 2\eps)} 
  \sigma_{ab}^{(0)}(M^2) 
\\ &\quad 
  \times  \bigg[
    \frac{{\cal A}_{-2}}{\eps^2}  +  \frac{{\cal A}_{-1}}{\eps}  +  {\cal A}_0 \! \bigg]
+  {\cal O}(\eps)\ .
\esp\ee
It should be noted that for supersymmetric processes, these virtual corrections not only
include pure QCD contributions, but also supersymmetric diagrams with squarks and gluinos
running into the loops.

In order to avoid a double-counting of the logarithmic terms when combining the resummed
cross section with the fixed-order result as given in Eq.~\eqref{eq:mtch}, it is necessary to
expand Eq.~\eqref{eq:resth} at ${\cal O}(\alpha_s)$. The expanded partonic cross section is
then given, in Mellin space, by \cite{Bozzi:2007qr,Debove:2010kf}
\be\bsp
  &\ \sigma_{ab}^{\rm(exp.)}(N,M^2,\mu_F^2, \mu_R^2) =
    \tilde{\cal H}_{ab}^{(0)}(M^2,\mu^2_R) \\
  &\ \ +
    \frac{\alpha_s}{2 \pi} \tilde{\cal H}_{ab}^{(1)}(M^2,\mu^2_R) \\
  &\ \ -\frac{\alpha_s}{2\pi} \ln\frac{\bar{N}^2 \mu_F^2}{M^2} \sum_{c}
    \Big[\gamma_{ac}^{(1)}(N) \tilde{\cal H}_{cb}^{(0)}(M^2,\mu_R^2)
     \Big] \\
  &\ \ -\frac{\alpha_s}{2\pi} \ln\frac{\bar{N}^2 \mu_F^2}{M^2} \sum_{c}
    \Big[
       \tilde{\cal H}_{ac}^{(0)}(M^2,\mu_R^2) \gamma_{bc}^{(1)}(N)
     \Big] \\
  &\ \ -\frac{\alpha_s}{2 \pi} \tilde{\cal H}_{ab}^{(0)}(M^2,\mu_R^2) 
    \Big[(A_a^{(1)}+A_b^{(1)}) \ln^2\bar{N} \Big] \\
  &\ \ +\frac{\alpha_s}{\pi} \tilde{\cal H}_{ab}^{(0)}(M^2,\mu_R^2) 
    \Big[ (\gamma_a^{(1)}+\gamma_b^{(1)})  \ln\bar{N} \Big] \ ,
\esp\ee
the quantities $\gamma_{ab}^{(1)}$ being the full Mellin moments of the one-loop approximation
of the Altarelli-Parisi splitting functions in four dimensions,
\be\bsp
 \gamma_{qq}^{(1)}(N) =&\ C_F\Big[\frac{3}{2}+\frac{1}{N(N+1)}-2\sum_{k=1}^N\frac{1}{k}\Big] \ , \\
 \gamma_{gq}^{(1)}(N) =&\ C_F\Big[\frac{2+N+N^2}{N(N^2-1)}\Big]\ ,\\
 \gamma_{qg}^{(1)}(N) =&\ \tau_R\Big[\frac{2+N+N^2}{N(N+1)(N+2)} \Big] \ ,\\
 \gamma_{gg}^{(1)}(N) =&\ 2C_A \! \Big[\frac{1}{N(N\!-\!1)} \!+\! \frac{1}{(N\!+\!1)(N\!+\!2)}
    \!-\! \sum_{k=1}^N\! \frac{1}{k}\Big] \!\!+\! \beta_0 \ .
\esp\label{eq:AP}\ee
and $\gamma_a^{(1)}$ the field anomalous dimensions corresponding, in axial gauge, to the
virtual, $N$-independent, pieces of $\gamma_{aa}^{(1)}$~\cite{Frazer:1978jd}, 
\be
 \gamma_q^{(1)} = {3C_F\over2} \quad\text{and}\quad \gamma_g^{(1)} = \beta_0 \ .
\label{eq:ano}\ee

\subsection{Transverse-momentum resummation at the next-to-leading logarithmic accuracy}
\label{sec:pt}
In order to be able to refactorize Eq.~\eqref{eq:d2sig} without performing the integration over
the transverse momentum $p_T$ and hence subsequently resum the large logarithmic contributions
arising at small $p_T$, it is necessary to apply a Fourier transform to the partonic cross
section $\sigma_{ab}$,
\be\bsp
 & \sigma_{ab}^{\rm (res.)}(N,M^2,p_T^2,\mu_F^2, \mu^2_R) =
   \int_0^\infty \d b\frac{b}{2}\ J_0(bp_T)\ 
\\ &\qquad \qquad \times
   \sigma^{\rm (res.)}_{ab}(N,M^2,b^2,\mu_F^2,\mu_R^2) \ ,
\esp\label{eq:fourier}\ee
where $J_0(y)$ denotes the zeroth-order Bessel function. This operation renders the cross
section explicitly dependent on the variable $b$, conjugate to the
transverse momentum $p_T$, dubbed the impact parameter. This additional transformation
allows us to correctly take into account transverse-momentum conservation so that
the partonic cross section can be rewritten under a form where soft and collinear radiation
exponentiates~\cite{Collins:1981va,Collins:1981uk,Collins:1984kg},
\be\bsp
 &\ \sigma_{ab}^{\rm (res.)} (N,M^2,b^2,\mu_F^2, \mu_R^2) = \sum_{a',\tilde a,b',\tilde b}
   E_{a'a}^{(1)}(N,1/\bar{b}^2,\mu_F^2) \\
&\ \times
   E_{b'b}^{(1)}(N,1/\bar{b}^2,\mu_F^2) \
   {\cal C}_{\tilde aa'}(N, 1/\bar{b}^2) \
   {\cal C}_{\tilde bb'}(N, 1/\bar{b}^2) \\
 &\ \ \times
   {\cal H}_{a''b''}(M^2, \mu_R^2)\ 
   \exp\Big[{\cal G}_{a''b''}(M^2 \bar{b}^2,M^2,\mu_R^2)
    \Big] \ .
\esp \label{eq:respt}\ee
In this equation, holding at the next-to-leading logarithmic accuracy,
the presence of the one-loop approximation of the QCD evolution operators allows for
evaluating the parton densities at the
natural scale of the process $1/\bar{b}$, with $\bar b \equiv (b/2) e^{\gamma_E}$.
Moreover, all the other functions can be calculated perturbatively.

Although there
are some freedoms, corresponding to the choice of a resummation scheme, in the way to separate
the different contributions into the various ${\cal C}_{ab}$, ${\cal G}_{ab}$ and
${\cal H}_{ab}$ factors \cite{Catani:2000vq,Bozzi:2005wk}, we adopt the
most physical option where the Sudakov form factor and the ${\cal C}_{ab}$ function are free
from any hard contribution. In this case, the Sudakov form factor is written
as~\cite{Bozzi:2006fw,Debove:2009ia}
\be\bsp
  {\cal G}_{ab}(M^2 \bar{b}^2,M^2,\mu_R^2) = &\ g_{ab}^{(1)} \Big(\frac{\alpha_s}{2 \pi}
    \beta_0 \ln[M^2\bar{b}^2]\Big) \ln[M^2\bar{b}^2]
\\ &\qquad  +
    g_{ab}^{(2)} \Big(\frac{\alpha_s}{2 \pi} \beta_0 \ln[M^2\bar{b}],
   \frac{M^2}{\mu_R^2} \Big) \ ,
\esp\ee
where the first term in this expansion,
\be\label{eq:ptg1}
 g_{ab}^{(1)}(\lambda) = \frac{1}{2\lambda\beta_0}
   (A_a^{(1)}+A_b^{(1)}) \big[\lambda+\ln(1-\lambda)\big] \ ,
\ee
collects the leading logarithmic contributions, and the second term,
\be\label{eq:ptg2}\bsp
 & g_{ab}^{(2)}(\lambda,M^2/\mu^2_R) =
  \frac{1}{2\beta_0} \Big[B_a^{(1)}+B_b^{(1)}\Big]  \ln(1-\lambda)  \\
 &\ + \frac{1}{2\beta_0} \Big[A_a^{(1)}+A_b^{(1)}\Big]\Big[\frac{\lambda}{1-\lambda} +
      \ln(1-\lambda)\Big] \ln\frac{M^2}{\mu_R^2} \\
 &\ + \frac{\beta_1}{2 \beta_0^3} \Big[A_a^{(1)}\!+\!A_b^{(1)}\Big] \Big[
  \frac{\lambda+\ln(1-\lambda)}{1-\lambda}
  +\frac{1}{2}\ln^2(1-\lambda)\Big] \\
 &\ - \frac{1}{2 \beta_0^2} \Big[A_a^{(2)}+A_b^{(2)}\Big] \Big[\frac{\lambda}{1-\lambda}
  +\ln(1-\lambda)\Big] \ ,
\esp\ee
the next-to-leading pieces. We recall that the relevant coefficients of the
resummation functions $A_a$ and $B_a$ have already been introduced in Eq.\ \eqref{eq:A1A2} and
Eq.~\eqref{eq:B1}.

In the `physical' resummation scheme that we have adopted, the hard function ${\cal H}_{ab}$ is
free from any
logarithmic contribution and includes, as for threshold resummation, the finite parts of the
renormalized virtual contributions ${\cal A}_0$ defined in Eq.~\eqref{eq:defa0}. It reads, at
the next-to-leading order accuracy~\cite{Bozzi:2006fw,Debove:2009ia},
\be
  {\cal H}_{ab}(M^2,\mu_R^2) = \sigma_{ab}^{(0)}(M^2) \Big[ 1 + \frac{\alpha_s}{2 \pi}
     {\cal A}_0\Big] \ .
\label{eq:ptH}\ee
Finally, the ${\cal C}_{ab}$ functions are evaluated, still at the next-to-leading logarithmic
accuracy and in this scheme, as \cite{Bozzi:2006fw,Debove:2009ia}
\be\label{eq:ptC}
  {\cal C}_{ab}(N,\mu_R^2) = \delta_{ab} + \frac{\alpha_s}{2 \pi} \Big[\frac{\pi^2}{6} C_a 
     \delta_{ab}-\gamma_{ab}^{(1),\eps}(N) \Big] \ ,
\ee
where $\gamma_{ab}^{(1),\eps}$ denotes the ${\cal O}(\eps)$ parts of the Altarelli-Parisi
splitting kernels in Mellin space,
\be\bsp
  \gamma_{qq}^{(1),\eps}(N) =&\ \frac{-C_F}{N(N+1)}\  ,\\
  \gamma_{qg}^{(1),\eps}(N) =&\ \frac{-2\tau_R}{(N+1)(N+2)},\\
  \gamma_{gq}^{(1),\eps}(N) =&\ \frac{-C_F}{N+1}\ , \\
  \gamma_{gg}^{(1),\eps}(N) =&\ 0 \ .
\esp \ee

After resumming the partonic cross section in the impact parameter
$b$-space, the resummed cross section has to be transformed back to the physical $p_T$-space.
This procedure requires to pay a particular attention to the singularities present in the
resummed exponent when $\lambda=1$ in Eq.~\eqref{eq:ptg1} and Eq.~\eqref{eq:ptg2} that
are related to the presence of the Landau pole in the perturbative running of the strong
coupling constant. Following the prescription presented in Ref.~\cite{Laenen:2000de}, the
inverse Fourier transform is calculated after deforming the integration contour of the
$b$-integral into the complex plane by defining two integration branches
\be
  b = e^{\pm i\varphi}\ t \quad\text{with}\quad t\in[0,\infty[ \quad\text{and}\quad
   \varphi \in ]0, \frac{\pi}{2}[ \ .
\label{eq:ctr}\ee
The Bessel function $J_0(y)$ appearing in Eq.~\eqref{eq:fourier} is then replaced by the sum
of two auxiliary functions $h_1$ and $h_2$ that distinguish positive and negative phases in the
complex $b$-plane,
\be\bsp
  h_1(y,v) = & -\frac{1}{2\pi}\int_{-iv\pi}^{-\pi+iv\pi}\d\theta\ e^{-iy\sin\theta}\ ,\\
  h_2(y,v) = & -\frac{1}{2\pi}\int^{-iv\pi}_{\pi+iv\pi}\d\theta\ e^{-iy\sin\theta}\ .
\esp\ee
For any choice of the $v$-parameter, these two functions are always finite and their sum is
independent of $v$. This splitting has the advantage that each of the two functions is
associated with only one single branch of the integration contour of Eq.~\eqref{eq:ctr}.

In order to match with the fixed order result, making use of Eq.~\eqref{eq:mtch}, the resummed
cross section of Eq.~\eqref{eq:fourier}, together with Eq.~\eqref{eq:respt},
we expand these two equations at order~${\cal O}(\alpha_s)$,
\be\bsp
 & \sigma_{ab}^{\rm(exp.)}(N,M^2,p_T^2,\mu_F^2,\mu_R^2) =
 {\cal H}_{ab}^{(0)}(M^2,\mu^2) \\
 &\ + \frac{\alpha_s}{2 \pi} {\cal H}_{ab}^{(1)}(M^2,\mu^2) \\
 &\ - \frac{\alpha_s}{2 \pi} \Big[ 2{\cal J} \!-\! \ln \frac{M^2}{\mu^2_F}\Big] \sum_c
   \Big[{\cal H}_{ac}^{(0)}(M^2,\mu_R^2) \gamma_{cb}^{(1)}(N)
     \Big] \\
 &\ -\! \frac{\alpha_s}{2 \pi} \Big[ 2{\cal J} \!-\! \ln \frac{M^2}{\mu^2_F}\Big] \sum_c
   \Big[\gamma_{ca}^{(1)}(N) {\cal H}_{cb}^{(0)}(M^2,\mu_R^2)\Big] \\
 &\ + \frac{\alpha_s}{2\pi} \sum_c
   \Big[{\cal H}_{ac}^{(0)}(M^2,\mu_R^2) {\cal C}_{cb}^{(1)}(N) \Big] \\
 &\ + \frac{\alpha_s}{2\pi} \sum_c
   \Big[ {\cal C}_{ca}^{(1)}(N) {\cal H}_{cb}^{(0)}(M^2,\mu^2_R)\Big] \\
 &\ - \frac{\alpha_s}{8 \pi} {\cal H}_{ab}^{(0)}(M^2,\mu_R^2)
    \Big[A_a^{(1)}+A_b^{(1)}i\Big] {\cal J}^2 \\
 &\ - \frac{\alpha_s}{4 \pi} {\cal H}_{ab}^{(0)}(M^2,\mu_R^2)
    \Big[B_a^{(1)}+B_b^{(1)}\Big]{\cal J}  \ .
\esp\label{eq:ptexp}\ee
We recall that the resummation coefficients appearing in this expression have already been
introduced in
Eq.~\eqref{eq:A1A2} and Eq.~\eqref{eq:B1} and that the Altarelli-Parisi splitting kernels have
been presented in Eq.~\eqref{eq:AP}. Moreover, the first two
coefficient of the perturbative expansion of the hard function ${\cal H}_{ab}$ and those of the
function ${\cal C}_{ab}$ are deduced from Eq.~\eqref{eq:ptH} and Eq.~\eqref{eq:ptC}. In
addition, all the dependence on the transverse momentum has been embedded within the integral 
$\cal J$ defined by
\be
  {\cal J} = \int_0^\infty \d b\ \frac{b}{2}\ J_0(bp_T)\ \ln[M^2 \bar{b}^2] \ .
\ee

\subsection{Joint resummation at the next-to-leading logarithmic accuracy}
\label{sec:joint}
In this section, we generalize the results of Section \ref{sec:pt} so that both types of large
logarithms arising either in the small $p_T$ region or near threshold are resummed
simultaneously. Since these logarithms have the same dynamical origin, their joint
reorganization is possible. In this way, they eventually exponentiate very similarly to
the case of the transverse momentum regime of Eq.~\eqref{eq:respt}~\cite{Laenen:2000de,%
Li:1998is,Laenen:2000ij},
\be\bsp
 & \sigma_{ab}^{\rm (res.)} (N,M^2,b^2,\mu_F^2, \mu_R^2) = \sum_{a',\tilde a,b',\tilde b}
   E_{a'a}^{(1)}(N,M^2/\chi^2,\mu_F^2) \\
 &\ \times
   E_{b'b}^{(1)}(N,M^2/\chi^2,\mu_F^2) 
   {\cal C}_{\tilde aa'}(N, M^2/\chi^2)
   {\cal C}_{\tilde bb'}(N, M^2/\chi^2)
\\ &\  \times
   {\cal H}_{a''b''}(M^2, \mu_R^2)
   \exp\Big[{\cal G}_{a''b''}(M^2, \bar{N}, \bar{b},\mu_R^2) \Big] \ .
\esp\label{eq:joint}\ee
In order to ensure a proper refactorization of the cross section, a Fourier transform has again
been performed, as in Eq.~\eqref{eq:fourier}. Furthermore, we have introduced the function
$\chi$, defined by
\be
 \chi\equiv \chi(\bar{N}, \bar{b}) = \frac{\bar{N}}{1 + \bar{b}/\bar{N}}+\bar{b} \ ,
\label{eq:chi}\ee
which interpolates between $\bar{N}$ in the threshold region, when $\bar{N}\gg\bar{b}$,
and $\bar{b}$ in the small-$p_T$ region, when $\bar{b}\gg\bar{N}$. Even though there are
several ways to define such an interpolation, the choice of Eq.~\eqref{eq:chi} first implies
that the leading and next-to-leading logarithms both in $\bar{b}$ and $\bar{N}$ are correctly
reproduced, respectively in the limits $\bar{b}\to \infty$ and $\bar{N}\to\infty$. Next, it
avoids the introduction of sizable subleading terms into perturbative expansions in $\alpha_s$
of the resummed formula of Eq.~\eqref{eq:joint} that are not predicted by fixed-order
computations.

While the ${\cal H}_{ab}$ and ${\cal C}_{ab}$ functions have exactly the same form as their
counterparts in the small transverse-momentum regime shown in Eq.~\eqref{eq:ptH} and
Eq.~\eqref{eq:ptC}, the Sudakov form factor now reads
\be\bsp
  {\cal G}_{ab}(M^2,\bar{N}, \bar{b}^2,\mu_R^2) = &\ g_{ab}^{(1)} \Big(\frac{\alpha_s}{2 \pi}
    \beta_0 \ln\chi\Big) \ln\chi
\\ &\quad
    + g_{ab}^{(2)} \Big(\frac{\alpha_s}{2\pi}\beta_0\ln\chi,\ln\bar{N},\frac{M^2}{\mu_R^2}
     \Big) \ ,
\esp\ee
where the coefficients of its next-to-leading logarithmic accurate expansion are, in the
$\overline{\rm MS}$-scheme, given by~\cite{Bozzi:2007tea,Debove:2011xj}
\be\bsp
 & g_{ab}^{(1)}(\lambda)= \frac{1}{2\lambda\beta_0} (A_a^{(1)}+A_b^{(1)})
   \Big[2\lambda+\ln(1-2\lambda)\Big]\ , \\
 & g_{ab}^{(2)}(\lambda,\ln\bar{N},\frac{M^2}{\mu_R^2})=
  - \frac{1}{\beta_0} \Big[\gamma_a^{(1)} + \gamma_b^{(1)}\Big]\ln(1-2\lambda) 
\\&\
  - \! \frac{1}{2 \beta_0^2}\Big[A_a^{(2)} \!+\! A_b^{(2)}\Big]
   \Big[2\lambda\frac{1-\frac{\alpha_s}{\pi}\beta_0\ln\bar{N}}{1-2\lambda}
    + \ln(1-2\lambda)\Big]\\
 &\ +\! \frac{1}{\beta_0}\Big[A_a^{(1)} \!\!+\! A_b^{(1)}\Big] 
     \Big[\lambda\frac{1\!-\!\frac{\alpha_s}{\pi}
    \beta_0\ln\bar{N}}{1\!-\!2\lambda} \!+\! \frac12
     \ln(1\!-\!2\lambda)\Big]\ln\frac{M^2}{\mu_R^2} \\
 &\ +\! \frac{\beta_1}{2 \beta_0^3} \Big[A_a^{(1)} \!+\! A_b^{(1)}\Big] \Big[
  \frac{\big(2\lambda+\ln(1 \!-\! 2\lambda)\big)\big(1 \!-\! 
    \frac{\alpha_s}{\pi}\beta_0\ln\bar{N}\big)}{1\!-\!2\lambda} \Big] \\
 &\ +\! \frac{\beta_1}{4 \beta_0^3} \Big[A_a^{(1)} \!+\! A_b^{(1)}\Big] \Big[
     \ln^2(1\!-\!2\lambda)\Big] \  .
\esp\ee
We recall that the resummation coefficients have already been shown in Eq.~\eqref{eq:A1A2} and
Eq.~\eqref{eq:B1} and that the one-loop approximation of the field anomalous
dimensions $\gamma_a^{(1)}$ are the $N$-independent parts of the Altarelli-Parisi splitting
kernels (in axial gauge) given in Eq.~\eqref{eq:ano}.

Double counting implied when combining the resummed results presented above, after getting back
to the physical $p_T$-space as shown in Section \ref{sec:pt}, is again removed by subtracting
the expansion of Eq.\ \eqref{eq:fourier}, together with Eq.~\eqref{eq:joint}, at the first
order in $\alpha_s$. This expansion has the same functional form as Eq. \eqref{eq:ptexp},
after replacing the integral of the zeroth-order Bessel function ${\cal J}$ by
\be
  \tilde{\cal J} = \int_0^\infty \d b\ \frac{b}{2}\ J_0(bp_T)\ \ln\chi \ .
\ee

\section{Installing and running \resum}\label{sec:resum}
\subsection{Requirements and technical details}\label{sec:prereq}

In order to use \resum, several external libraries and header files are required and must be
installed on the system.

First, information on the benchmark supersymmetric scenario under consideration is passed to
the program by means of files compliant with the Supersymmetry Les Houches Accord
(SLHA) conventions~\cite{Skands:2003cj,Allanach:2008qq}. We have adopted the choice to
internally
handle such files by making use of \slhaea~\cite{slhaea}, a \cpp\ header-only library
dedicated to input, output and manipulation of SLHA data. While \slhaea\ is fully included
in \resum\ and hence does not need to be downloaded by the user, this tool relies on some
headers of the \boost\ \cpp\ libraries \cite{boost} that are in contrast not provided
with \resum. Therefore, their presence on the system is a necessary prerequisite and the two
packages {\sc boost} and {\sc boost-devel} must be available.

Next, both the fixed order and resummed components of the hadronic cross section require the
evaluation of parton distribution functions, either in the physical $x$-space or in the
conjugate Mellin $N$-space. The \resum\ package does not come with any built-in parton
density fit and entirely relies on the external \lhapdf\ library \cite{Whalley:2005nh} which
must therefore be installed by the user. On run-time, the Mellin moment of the
parton densities parameterization under consideration are obtained by a numerical fit performed
by means of the Levenberg-Marquardt algorithm dedicated to multidimensional fits of non-linear
functions~\cite{LM1,LM2}, as implemented in the {\sc Gnu} Scientific \cpp\ Libraries (\gsl). This
algorithm consists of an iterative procedure using the method of least squares after
a linearization of the fitting curves. As
a consequence, both \gsl\ header and library files must be installed by the user before being
able to run \resum.

The knowledge of the parton densities both in $x$-space and $N$-space allows to
compute all three components of the hadronic differential cross section $\d^2\sigma /
\d p_T \d M$ associated with the implemented physics processes, as described in
Section~\ref{sec:formalism}. Additional
integration upon the invariant mass $M$ of the gaugino or slepton pair or upon
their transverse momentum $p_T$ then leads to the singly differential cross
sections $\d\sigma/\d p_T$ and $\d\sigma/\d M$, respectively. Furthermore, integration upon
both variables
allows to extract total production rates. Let us note that in the case of threshold
resummation, the integration upon $p_T$ has been performed analytically and there is no way to
access doubly-differential cross sections.
All these integrations, together with the usual two-body and
three-body phase space integration relevant for the types of computations performed in \resum,
are achieved by means of an adaptive multi-dimensional integration technique based on the
importance sampling of the integration domain~\cite{veg1}. To this aim, we again make use of
the \gsl\ \cpp\ libraries provided with the adaptive multi-dimensional integration {\sc Vegas}
algorithm \cite{veg2}.

Concerning the fixed order partonic cross sections, \resum\ is based on the leading order
results of Ref.~\cite{Bozzi:2004qq} and Ref.~\cite{Bozzi:2007me,Debove:2008nr} for slepton-pair
and gaugino-pair production, respectively. Next-to-leading order corrections including both the
QCD and supersymmetric QCD virtual diagrams are implemented as given in
Ref.~\cite{Bozzi:2007qr} and Ref.~\cite{Debove:2010kf}, the associated finite pieces of the
virtual loops being computed by means of the {\sc QcdLoop} package \cite{Ellis:2007qk}. Since
the latter is fully embedded within \resum\ and thus does not need to be installed by the user.

\subsection{Installation}
We recommend the user to always use the latest stable version of \resum\ that can be
downloaded from the webpage

\noindent \verb+http://www.resummino.org+

\noindent Once downloaded, the package consists of a compressed tar file
(\texttt{resummino-x.x.x.tar.bz2} where \texttt{x-x-x} stands for the version number) that must
be unpacked,
\begin{verbatim}
tar xf resummino-x.x.x.tar.bz2
\end{verbatim}
In the case all the prerequisite dependencies of \resum\ are present on the system (see
Section~\ref{sec:prereq}), it is then necessary to generate a \texttt{Makefile} appropriate to
the system configuration. This is done by issuing in a shell the commands
\begin{verbatim}
cd resummino-x.x.x
cmake . [options]
\end{verbatim}

The \texttt{cmake} program checks, in a first stage, that all the dependencies mandatory for
\resum\ are correctly installed. It subsequently creates a series of
\texttt{Makefile} scripts allowing for the
compilation of the \resum\ source files and their linking with the dependencies. Two optional
arguments can be passed to the \texttt{cmake} script. The first of these is related
to the \lhapdf\ libraries. In the case they have not been installed in the directories
referred to by the environment variable \texttt{LD\textunderscore LIBRARY\textunderscore PATH}
(or \texttt{DYLD\textunderscore LIBRARY\textunderscore PATH} for {\sc MacOS} systems), the
\lhapdf\ installation directory must be specified by means of
\begin{verbatim}
-DLHAPDF=/path/to/lhapdf
\end{verbatim}
This instructs \texttt{cmake} that the \lhapdf\ libraries are stored in the directory
\texttt{/path/to/lhapdf/lib} and the header files in the directory
\texttt{/path/to/lhapdf/include}. Equivalently, these two directories can be provided
separately through the \texttt{cmake} options
\begin{verbatim}
-DLHAPDF_LIB_DIR=/path/to/lhapdf/lib 
-DLHAPDF_INCLUDE_DIR=/path/to/lhapdf/include
\end{verbatim}

The second optional argument of the \texttt{cmake} script consists of information on the
directory where the \resum\ executable has to be created (\texttt{/path/to/install} in the
example below). This is specified by including the option
\begin{verbatim}
-DCMAKE_INSTALL_PREFIX=/path/to/install
\end{verbatim}
when issuing the \texttt{cmake} command.

The \texttt{Makefile} can eventually be executed in order to generate a local release of \resum
\begin{verbatim}
make
make install
\end{verbatim}
which can then be further used for physics applications.

\subsection{Running the code}\label{sec:runn}
Once compiled, \resum\ can be immediately run from a shell by issuing
\begin{verbatim}
resummino filename
\end{verbatim}
where \texttt{filename} consists of the path to a file containing the settings of the
calculation to be performed.  Three extra modes of running can also be employed by adding an
optional flag when executing the code,
\begin{verbatim}
resummino --lo filename
resummino --nlo filename
resummino --parameter-log=params.log filename
\end{verbatim}
The first two choices above allow to respectively compute leading-order
and next-to-leading order quantities (without ma\-tching to a resummation calculation).
In contrast, the last of the three options leads to the generation of a file denoted by
\texttt{params.log} that includes all the numerical values of the parameters defining the
supersymmetric benchmark scenario under consideration.

Now, we turn to the way to encode the computation information in the input file to be parsed
when executing the code. First, the definition of the
collider is passed to the program by fixing the
nature of the colliding beams and the hadronic center-of-mass energy (to be given in GeV). This
is achieved by configuring the variables \texttt{collider\textunderscore type} and
\texttt{center\textunderscore of\textunderscore mass\textunderscore energy}, \ie, by including
in the input file lines of the form of
\begin{verbatim}
collider_type = proton-proton
center_of_mass_energy = 8000
\end{verbatim}
Let us note that in the current version of the program, only proton-proton and
proton-antiproton collisions are supported, so that the variable
\texttt{collider\textunderscore type} can only be set to one of the values
\texttt{proton-proton} and \texttt{proton-antiproton}.

Next, the two produced superparticles must
be referred to by setting the variables \texttt{particle1} and \texttt{particle2} to the
relevant Particle Data Group (\texttt{PDG}) codes \cite{Beringer:1900zz}. We recall that
since \resum\ is strictly dedicated to the production of the electroweak superpartners, only
slepton, sneutrino, chargino and neutralino states are allowed as final state particles. For
instance, the production of a lightest neutralino (whose the PDG code is 1000022) in
association with a negatively-charged
next-to-lightest chargino (whose the PDG code is -1000037) is encoded as
\begin{verbatim}
particle1 =  1000022
particle2 = -1000037
\end{verbatim}
The numerical values of the masses of those particles, together with these of all the other
supersymmetric model parameters, are provided by means of a file compliant with the SLHA
conventions~\cite{Skands:2003cj,Allanach:2008qq}, as
already mentioned in Section~\ref{sec:prereq}. The path to this file is specified as the
value of the variable \texttt{slha},
\begin{verbatim}
slha = slha.in
\end{verbatim}
where in the example above, \texttt{slha.in} denotes a generic SLHA file.

The last pieces of information to be included in the input file define the type of
computation to be performed and the numerical precision to be reached. The variable
\texttt{result} allows to select the observable to
compute by setting its value to \texttt{total} (total cross section $\sigma$ using
the threshold-resummation formalism), \texttt{pt}
(transverse-momentum distribution $\d\sigma/\d p_T$ using the $p_T$-resummation formalism),
\texttt{ptj}
(transverse-momentum distribution $\d\sigma/\d p_T$ using the joint-resummation formalism) or
\texttt{m} (inv\-a\-ri\-ant-mass
distribution $\d\sigma/\d M$ using the threshold-re\-sum\-ma\-ti\-on formalism).
For the last three possibilities, the numerical value of the
transverse-momentum and the one of the invariant-mass at which the differential cross section
must be respectively evaluated have to be referred to via the variables \texttt{pt} and
\texttt{M}. For instance, implementing in the input file
\begin{verbatim}
result = pt
pt = 50
\end{verbatim}
leads to the evaluation of $\d\sigma/\d p_T$ for $p_T = 50$ GeV, while
\begin{verbatim}
result = M
M = 600
\end{verbatim}
implies the evaluation of $\d\sigma/\d M$ for $M = 600$ GeV. Furthermore, having instead
\begin{verbatim}
result = total
\end{verbatim}
defines the computation of the total cross section. The parton density sets to be employed
for both the leading-order and higher-order components of the calculated observable are
indicated following the \lhapdf\ conventions which are based on an ordering
according to parton density group names and numbers~\cite{Whalley:2005nh}. For instance, the
command lines
\begin{verbatim}
pdf_lo     = MSTW2008lo68cl
pdfset_lo  = 0
pdf_nlo    = MSTW2008nlo68cl
pdfset_nlo = 0
\end{verbatim}
instruct \resum\ to use the best fits (the variables \texttt{pdfset\textunderscore lo} and
\texttt{pdfset\textunderscore nlo} are set to zero) of the leading-order and
next-to-leading-order fits of the MSTW 2008 parton densities~\cite{Martin:2009iq} (as indicated
by the variables \texttt{pdf\textunderscore lo} and \texttt{pdf\textunderscore nlo}).
Factorization and renormalization scales are internally
set to the sum of the mass of the produced particles,
up to additional factors that must be specified in the input file via the intuitive variables
\texttt{mu\textunderscore f} and \texttt{mu\textunderscore r}. For the sake of the example,
central scale choices where scales are fixed to the average mass of the produced particles
are enforced by
\begin{verbatim}
mu_f = 0.5
mu_r = 0.5
\end{verbatim}
Finally, the speed of the computation is driven by two parameters, the numerical precision to
be reached
and the maximum of iterations allowed when {\sc Vegas} is numerically computing the various
integrals presented in Section~\ref{sec:formalism}.
These are related to two input variables, \texttt{precision}, which takes a real number as a
value, and \texttt{max\textunderscore iter} which refers to an integer number. Hence, including
in the input file the lines
\begin{verbatim}
precision = 0.001
max_iters = 3
\end{verbatim}
allows for three iterations of {\sc Vegas} and demands a relative precision of~$0.1~\%$.

\section{Illustrative examples} \label{sec:benchmarking}
To illustrate the usage of \resum\ in a practical case, we perform
several calculations in the framework of one representative constrained scenario of the Minimal
Supersymmetric Standard Model (cMSSM). As designing an experimentally non-excluded
supersymmetric scenario is
going beyond the scope of this work, we refer to an earlier study performed by the LHC
Physics Center at CERN together with both the ATLAS and CMS supersymmetry working
groups~\cite{AbdusSalam:2011fc}. This analysis is based on 1 fb$^{-1}$ of LHC data, electroweak
precision observables and flavor physics constraints. Its conclusion consists of the proposal
of several reference points in the cMSSM parameter space to be used for supersymmetric searches
and phenomenological investigations.
We adopt their 31$^{\rm st}$ scenario, where the ratio
of the vacuum expectation values of the neutral components of the two Higgs doublets
$\tan\beta$ is set to 40 and the Higgs supersymmetric mixing parameter $\mu$ is taken positive.
At the su\-per\-sym\-me\-try-breaking scale, the universal scalar mass $m_0$ is fixed to
400~GeV, the
universal gaugino mass $m_{1/2}$ to 600~GeV and the universal trilinear coupling $A_0$ to
--500 GeV. After renormalization group running down to the electroweak scale, squarks and
gluino are found heavy, with masses of about 1.5~TeV, with the exception of
the lightest stop and sbottom states. The large left-right mixing inferred by the important
negative value of $A_0$ indeed lowers their masses to 940~GeV and 1100~GeV, respectively.
In contrast, all the other superpartners (neutralinos, charginos, sleptons and sneutrinos)
are lighter and lie in the 250--825 GeV range. In the following, we further restrict
ourselves to the production of the lightest electroweak superpartners whose masses are
approximately given by
\be\bsp
      m_{\tilde\chi_1^0}   = 250~\text{GeV}\ , \quad
  &\  m_{\tilde\chi_2^0}   =
      m_{\tilde\chi_1^\pm} = 480~\text{GeV}\ ,\\
      m_{\tilde e_L} =  m_{\tilde \mu_L} = 565~\text{GeV}\ , \quad
  &\  m_{\tilde e_R} =  m_{\tilde \mu_R} = 460~\text{GeV}\ ,\\
      m_{\tilde \tau_1} = 295~\text{GeV}\ , \quad
  &\  m_{\tilde \tau_2} = 535~\text{GeV}\ .
\esp\ee

%
\begin{table}[!t]
\caption{\label{tab:total}
 Total cross sections associated with the production of any pair of superpartners among the
 lightest gauginos and sleptons ($\tilde \ell$ equivalently denotes mass-degenerate selectrons
 and smuons) in the context of the
 LHC collider running at a center-of-mass energy of 8~TeV and for the benchmark
 scenario~31 of Ref.~\cite{AbdusSalam:2011fc}. The
 results are computed at the leading order (LO) and next-to-leading order (NLO) of
 perturbative QCD and then matched to threshold resummation at the next-to-leading logarithmic
 accuracy (NLL+NLO). The corresponding scale uncertainties are also indicated and resummed
 cross sections smaller than 0.05 fb are omitted. }
\renewcommand{\arraystretch}{1.3}
\begin{center}
\begin{tabular}{| c || l | l | l |}
\hline
 Final & \multirow{2}{*}{LO [fb]} & \multirow{2}{*}{NLO [fb]} &\multirow{2}{*}{NLO+NLL [fb]}\\
 state & & & \\
\hline
$\tilde\chi^0_1 \tilde\chi^0_1$ &  $0.1245^{+8.6\%}_{-7.5 \%}$  & $0.1605^{+3.6\%}_{-3.6\%}$
   & $0.1554^{+ 0.2\%}_{-0.0\%}$\\
$\tilde\chi^0_2 \tilde\chi^0_2$ &  $0.0875^{+12\%}_{-10\%}$  & $0.1065^{+4.5\%}_{-3.7\%}$
   & $0.1043^{+0.3\%}_{-0.0 \%}$\\
\hline
$\tilde\chi^+_1 \tilde\chi^0_2$ &  $4.3674^{+9.9\%}_{-8.5\%}$  & $4.8750^{+2.0\%}_{-2.4\%}$
   & $4.8248^{+0.3 \%}_{-0.5 \%}$\\
$\tilde\chi^-_1 \tilde\chi^0_2$ &  $1.4986^{+10\%}_{-8.6\%}$  & $1.7333^{+2.1\%}_{-2.4 \%}$
   & $1.7111^{+0.6\%}_{-1.1 \%}$\\
\hline
$\tilde\chi^+_1 \tilde\chi^-_1$ &  $2.8874^{+9.9\%}_{-8.5\%}$  & $3.3463^{+3.3\%}_{-3.3\%}$
   & $3.3086^{+0.7\%}_{-0.3 \%}$\\
\hline
$\tilde \ell^+_R \tilde \ell_R^-$ & $0.0749^{+11\%}_{-9.1 \%}$ & $0.0868^{+2.7\%}_{-3.0\%}$
   & $0.0854^{+0.2\%}_{-0.4\%}$\\
$\tilde\ell^+_L \tilde\ell_L^-$ & $0.0477^{+12\%}_{-10\%}$ & $0.0543^{+2.8\%}_{-3.4\%}$
   & $0.0534^{+0.5\%}_{-0.3\%}$\\
$\tilde\tau^+_1\tilde\tau_1^-$ & $0.5878^{+7.6\%}_{-5.3\%}$ & $0.7093^{+2.5\%}_{-2.5\%}$
   & $0.6985^{+0.0\%}_{-0.2\%}$\\
\hline
\end{tabular}
\end{center}
\end{table}
%

Considering the LHC collider, running at a center-of-mass energy of 8~TeV, we
focus in Table~\ref{tab:total} on the largest total cross sections associated with
the production of any pair of two of the particles under
consideration. We indicate, in the second column of the table, results at the leading-order of
perturbative QCD,
employing the leading order set of the 2008 MSTW parton densities~\cite{Martin:2009iq}. In the
third column, we compute next-to-leading order predictions, convolving the
partonic cross section with the next-to-leading order set of the same parton
density fit. Finally, in the fourth column, these last results
are matched to threshold resummation. Although this does not imply a sensible change in the
cross sections, we emphasize the importance of resummation by showing
the theoretical uncertainties that are obtained after multiplying and
dividing the central scale, set to the average mass of the produced particles (see
Section~\ref{sec:runn}), by a factor of two.
Stabilization of the results can indeed be observed once soft and collinear radiation is
resummed to all orders in $\alpha_s$. At the leading-order
accuracy, the evolution of the parton densities introduces a
dependence on the factorization scale through potentially large logarithmic terms, which leads
to an uncertainty of about $\pm 10\%$. Although this specific source of
uncertainties is reduced at the next-to-leading
order, new ${\cal O}(\alpha_s)$ diagrams imply an additional dependence on the
renormalization scale. This yields a total scale uncertainty of a few percents.
Finally, exponentiation,
which allows to account for the dominant higher-order contributions within the Sudakov form
factor, permits to render scale variations under a very good control, at the percent level.

\begin{figure}
  \includegraphics[width=.90\columnwidth]{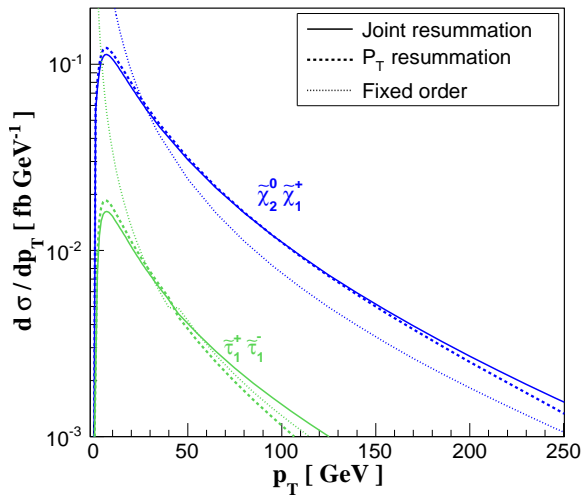}
\caption{Transverse-momentum distributions for $\tilde\chi_2^0\tilde\chi_1^+$ and 
  $\tilde\tau_1^+\tilde\tau^-_1$ production at the LHC,
  running at a center-of-mass energy of 8 TeV, at order ${\cal O}(\alpha_s)$ (dashed) and after
  matching the results with transverse-momentum (dotted) and joint (plain) resummation.}
\label{fig:pt}
\end{figure}

In Figure~\ref{fig:pt}, we present transverse-momentum spectra related to the production of a
selection of light
chargino, neutralino and slepton pairs at the LHC. We first compute the predictions at
${\cal O}(\alpha_s)$ (dashed) and we next match the results with resummation in the
transverse-momentum (dotted) and joint (plain) regimes.
While the fixed-order results diverge at small
transverse momentum due to the large logarithmic terms that have to be exponentiated, their
resummation leads to a finite (and physical) behavior with a pronounced peak in the
region where
$p_T\lesssim 10$ GeV. In this kinematical range, the asymptotic expansion of the two
resummation formulas are in good agreement with the ${\cal O}(\alpha_s)$ results since they are
all dominated by
the logarithms. Therefore, matching with resummation as presented in Eq.~\eqref{eq:mtch}
allows for the regularization of the next-to-leading order predictions
for small values of $p_T$. The same matching effects also imply that the resummed predictions
are sensibly larger than the fixed order ones when the transverse momentum of the superpartner
pair lies within the
intermediate $p_T$-range of 20--60~GeV. Finally, it is also shown that calculations
using transverse-momentum and joint resummation agrees reasonably well with each other on the
entire $p_T$-range, although based on different Sudakov form factors.

\section{Summary} \label{sec:summary}
In this paper, we have introduced the \resum\ package, a \cpp\ and \fortran\ program dedicated
to precision calculations for gaugino and slepton pair production at hadron colliders. The
program allows to compute total cross sections, invariant-mass and transverse-momentum
distributions at leading order and next-to-leading order of perturbative QCD. In addition, the
results are then matched to a resummation of the large logarithmic terms appearing at
fixed order according to the transverse-momentum, threshold or joint resummation formalism.

We have illustrated the usage of our code by
adopting a typical supersymmetric benchmark scenario for superpartner searches and
performing in this context various computations by means of \resum. In the presented selection
of results, we have chosen to emphasize the major advantages of making use of resummed
predictions, \ie, a drastic reduction of the associated scale uncertainties and
a regularization of the transverse-momentum spectrum in the small-$p_T$ region.

\section*{Acknowledgments}
This work has been supported by the BMBF
Theorie-Verbund and by the Theory-LHC-France initiative of the CNRS/IN2P3.

\bibliographystyle{spphys}
\bibliography{biblio}

\end{document}